# HIGH-ENERGY γ-RAY SOURCES AND THE QUEST FOR THEIR IDENTIFICATION

Patrizia A. Caraveo
Istituto di Astrofisica Spaziale e Fisica Cosmica del CNR
Via Bassini, 15
20133 Milano- ITALY (pat@mi.iasf.cnr.it)

## Abstract

The 3rd EGRET catalogue lists 271 sources and provides an identification for 101 of them. Thus, the unidentified sources are the majority in the γ-ray sky, as we currently know it. Solving the mystery of the unidentified γ-ray sources is a major challenge for all fields of astronomy. Here we shall review the different approaches towards source identification as well as the results achieved so far.

# Introduction

The 3rd EGRET catalogue (Hartman et al, 1999) lists 271 sources distributed all over the sky (see Figure 1). The source distribution is not uniform, showing a marked concentration towards the galactic plane.

The size of a typical γ-ray error box ( about 1 sq deg) prevents any straight identification of these sources. To overcome such limited angular resolution of our γ-ray telescopes, one has to rely on some peculiar source characteristics. Of these, the best example is a source time signature.

Indeed, virtually all the identifications achieved so far are based on either long-or short-term variability. Pulsation allows unambiguous association between radio pulsars and γ-ray sources, while correlated, short or long-term radio, optical or X-ray variability is the precious tool to recognize AGNs.

This is why all sources identified so far fall in one of these two classes: pulsating neutron stars and Active Galactic Nuclei. Although nobody believes that these two classes exhaust all possible γ-ray source scenarios, we have not succeeded, so far, to secure identification with other promising classes of sources

The squares in Figure 1 represent the 6 pulsars officially detected by EGRET, namely the classical radio pulsars in Crab, Vela , PSR 1706-44, PSR 1951+32, PSR 1055-58, together with the radio quiet Geminga. Please note that PSR 1951+32 is added for completion only, since it is just seen as a pulsed emission but it does not rank in the official list of 3EG sources. Indeed, the official GRO list of γ-ray emitting pulsars (Thompson et al, 2001) includes also PSR 1509-58, which is detected only up to 10 MeV (Kuiper et al, 1999). Few more isolated neutron stars, such as PSR B0656+14 (Ramanamuthy et al, 1996), PSR 1046-58 (Kaspi et al., 2000) and the msec pulsar PSR J0218+4232 (Kuiper et al.,2000 ) have tantalizing, but not yet conclusive, evidence of pulsation in high energy γ-rays.
The diamonds in Figure 1 indicate all the sources associated with radio loud AGNs, be they high confidence identifications (66 cases) or probable ones (27 occurrences).
Variability, sometimes on time scale of days, is a common features of these sources. Indeed, the detection of similar behaviour at other wavelengths is the key factor behind their identifications. The discovery of so many AGNs has been one of the major breakthroughs of the EGRET mission. Since the Galaxy is transparent to high energy γ-rays, several AGNs, largely invisible in the radio, optical or X-ray bands owing to galactic gas and dust absorption, are expected in the γ–ray sky at low galactic latitude.
The circles represent the UGOs (Unidentified Gamma Objects) which appear to be concentrated on the galactic disk, and, especially, towards its central regions.

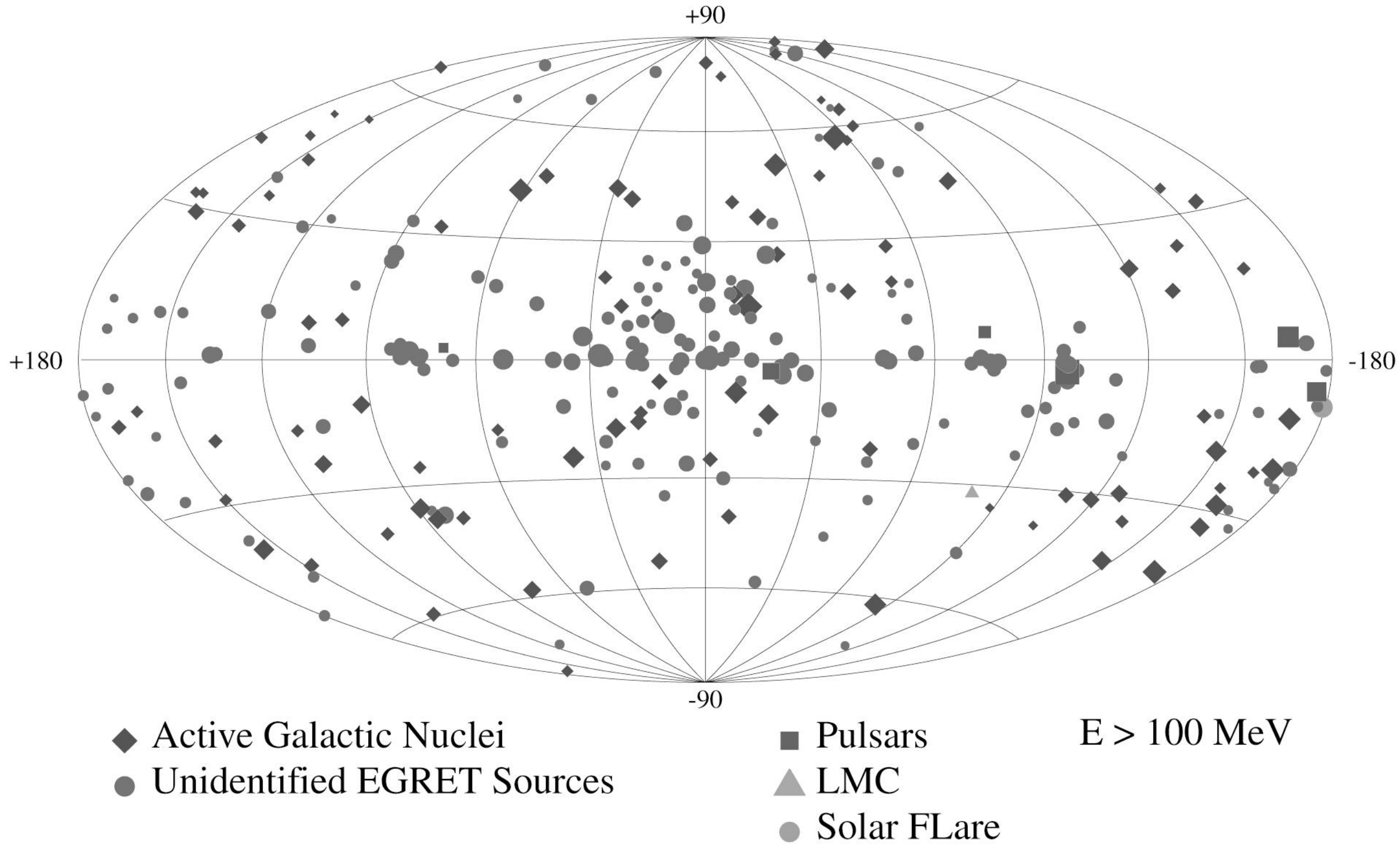

**Figure 1** The 3rd EGRET catalogue. The size of the symbols is proportional to the source flux.

## Exploiting the γ-ray data through a systematic approach

We know very little about the 170 unidentified EGRET sources. Apart from a rough location, the γ-ray instruments measure their fluxes and their spectra. Moreover, the study of the recorded count-rate allows one to assess their variability on time scales of days or weeks. Thus, source distribution, fluxes, spectra and count rates have been used as the only tools to constrain the nature of the unidentified sources

***Distribution, fluxes, spectra***

Studying source distribution as a function of galactic longitude and latitude provides constraints on sources' average distance and, thus, on their luminosity.
Using the source list available at the time (36 sources with b<10°), Mukherjee et al (1995) confirmed the early findings of the COS-B satellite (e.g. Bignami & Hermsen, 1983) and put

EGRET sources at distances between 1.2 and 6 kpc, implying isotropic luminosities in the range (0.7- 16.7) $10^{35}$ erg/sec.
The publication of the 3EG catalogue unveiled dozens of faint sources (small circles in Figure 1) at latitudes 10°<b<30°. This prompted Gehrels et al (2000) to redo the statistical study with the aim to compare the bright low latitude sources to the faint middle latitude ones. Gehrels et al (2000) have proposed this to be a new class of γ-ray sources, on average fainter and softer than the low latitude ones, possibly linked to the Gould's Belt structure (see also Grenier, 2000). Moreover, middle latitude sources should be closer to us than low latitude ones, thus their average luminosity should be lower.

***Variability***
Wallace et al. (2000) performed a systematic search for short-term variability of all EGRET sources and found that 6 catalogued sources do exhibit variability on timescales of 1 or 2 days. Four of these variable sources are unidentified, while the remaining two are PKS 0528+134 (already known to be variable in timescale of days, see Hunter et al., 1993) and the source 3EG J0222+4253 which could be associated with the BL Lac object 3C66A. In addition, Wallace et al. confirmed the variable behaviour of 3EG J0241-6103, tentatively associated with the binary system LSI 61.303 (see Kniffen et al, 1997 and Tavani et al, 1998) and 3EG J1837-0606, a strongly variable γ-ray source proposed by Tavani et al. (1997) as the prototype of a new class of variable, possibly galactic, emitters. Moreover, Wallace et al. also found two more transient sources not listed in the 3EG catalogue.
Since a γ-ray variability is of no use without a correlated variability at other wavelengths, these results call for more attention to the quick look analysis of future missions to allow the search for variable objects shortly after (or in parallel with) their detection in γ-ray.

## Struggling towards the identifications

Historically, the identifications of γ-ray sources has been pursued following two approaches : on the one side population studies, aimed at finding the classes of celestial objects characterized by distributions (and energetics) similar the that of γ-ray sources, and on the other a case-by-case approach, based on multiwavelength observations of individual sources.
Most unfortunately, population studies did not shed any light on the identification problem. Owing to their galactic distribution, low latitude γ-ray sources correlate well with most galactic tracers, such as star formation regions, O-B associations and SNRs, as originally suggested by Montmerle (1979) and, more recently, e.g. by Yadigaroglu and Romani (1997) and references therein. Inevitably, such statistical studies are of limited significance, since chance coincidences are always non negligible. As mentioned before, middle latitude sources seem to follow the shape of Gould's Belt. Although suggestive, this information is too vague to actually pinpoint their counterparts and to understand their nature.

***Pulsars***
Since pulsars are the only identified sources in our Galaxy, particular care was devoted to searching for correlations between radio pulsars and γ-ray sources, both on an individual basis and on a population point of view.
Nel et al (1996) examined 350 radio pulsars and found only few positional coincidences with γ-ray sources, and none of them could be confirmed through timing signatures.
Since then, a new radio survey, aimed at finding young pulsars, was carried out with the Parkes radio telescope, yielding hundreds of new pulsars (Manchester et al, 2001).

Using the Parkes data, D'Amico et al (2001) discovered two young, promising radio pulsars inside the error boxes of 3EG J1420-6038 and 3EG J1837-0606, while Camilo et al (2001) found the 21,000 y old PSR J1016-5957 inside the error box of 3EG 1013-5915.
In view of the timing noise, usually present in young pulsars, it will be difficult to search for the time signature of such objects in the EGRET data. They will be certainly studied by future $\gamma$-ray missions.
Torres, Butt and Camilo (2002) searched for positional coincidences between the low latitude EGRET sources and 368 new Parkes pulsars, taking into account also the pulsars' age, energetics and distance . Their efforts confirmed the early findings by D'Amico et al. (2001) and Camilo et al. (2001), but failed to yield any new plausible coincidence.
As already mentioned, tentative evidence for pulsation has been collected, so far, only from PSR B1046-58, proposed by Kaspi et al (2000) as the counterpart of 3EG J1048-5840, and from PSR B0656+14 (Ramanamuthy et al, 1996) as well as from the msec pulsar PSR J0218+4232 (Kuiper et al.,2000 ), although neither were ranked as 3EG sources. Once again, these will certainly be interesting targets for future $\gamma$-ray missions.

***Multiwavelength studies of individual sources***

X-rays are generally used to bridge the gap between the poor resolution achievable in $\gamma$-rays and the standards of optical or radio astronomy. Such an exercise is time consuming, since it usually involves several observing cycles with different instruments at different facilities, both in orbit and on the ground .
The identification of Geminga, schematically shown in figure 2, represents a good example of a successful 20-year long multiwavelength campaign.
As an established representative of the non-radio-loud INSs (as originally defined by Caraveo, Bignami and Trümper, 1996), Geminga offers an elusive template behaviour: prominent in high energy $\gamma$-rays, easily detectable in X-rays but downright faint in optical, with sporadic or no radio emission (see Bignami and Caraveo, 1996 for a review of the source multiwavelength phenomenology).
In spite of its successful identification, Geminga does not provide a viable template for the entire family of unidentified $\gamma$-ray sources. While its energetics ($L_\gamma$ = 3.3 $10^{34}$ erg/sec) meets typical requirements for middle latitude sources, presumably part of a more local galactic population (Gehrels et al. 2000), Geminga is certainly not adequate to account for the very low latitude (i.e. more distant) EGRET sources.
However, the strategy devised for the chase of Geminga seems to still be the best one to bridge the positional accuracy gap between $\gamma$-ray and optical astronomy.
Years of multiwavelength, world-wide efforts are now yielding the following candidate counterparts:

**3 Geminga-like**, radio quiet INSs:
quite appropriately, we start with 3EG J1835+5918, recently dubbed "the next Geminga" by Halpern et al (2002) (see also Mirabal and Halpern 2001; Reimer et al., 2001). It is the brightest unidentified $\gamma$-ray source: only 5 times fainter than Geminga in $\gamma$-rays, it is 40 times less intense in X-rays but undetectable at optical wavelengths. 3EG J0010+7309 (Brazier et al, 1998) and 3EG J2020+4026 (Brazier et al., 1996) presumably belong to the same class. Since both 3EGJ 1835+5918 and 3EG J0010+7309 are middle latitude sources, their energetic requirements are easily compatible with a Geminga-like identification.

**2 peculiar binary systems:**
the periodically variable radio source GT 0236+610/LSI 61.303 had been proposed as the counterpart of COS-B source 2CG 135+01 by Bignami et al. (1981). In spite of ad hoc searches for correlated variability between the radio and the $\gamma$-ray emission (e.g. Kniffen et al, 1997;Tavani et al., 1998), no conclusive proof has been brought forward for the identification of 3EG 0241+6103 with GT 0236+610 /LSI 61.303.

For 3EG J0634+0521, Kaaret et al. (1999) have suggested the identification with SAX J0635+0533, a binary system containing a compact object with a Be star companion. Recently, such a proposed identification has been strengthened by the discovery of a 34 msec pulsation (Kareet, Cusumano and Sacco, 2000), coupled with a high period derivative, pointing towards a young, energetic pulsar in a binary system. This is an absolute first in the panorama of known binary systems, making SAX J0635+0533 an interesting system "per se".

For both 3EG J0241+6103 and 3EG J0634+0521, the physics behind their γ-ray production would be that of particle acceleration at the shock created by the pulsar interaction with the thick Be-star wind, especially during periastron passage.

**1 energetic young radio pulsar**, discovered in X-rays, to be added to the pulsar list discussed above. A newly discovered, energetic, young, isolated pulsar, showing 51.6 msec X-ray and radio pulsations, has been proposed by Halpern et al (2001a) as the counterpart of 3EG J2227+6122. No γ-ray timing signature is yet available.

**1 "galactic" Blazar;**

3EG J2016+3657 has been identified with a Blazar behind the Galactic plane (Mukherjee et al, 2000; Halpern et al, 2001b). This extragalactic "contamination", deep in the galactic plane, is to be expected, in view of the isotropic distribution of Blazars, coupled with the negligible absorption suffered by γ-ray photons through the galactic plane.

**few pulsar nebulae;**

Roberts, Romani and Kawai (2001) list a number of pulsar nebulae which could be associated, mostly on positional grounds, with a sample of GeV sources specially selected by the authors.

Of interest is the case of the Kookaburra Nebula, in the error box of 3EG J 1420-6038, where also one of the new pulsars of D'Amico et al. (2001) is located. Oka et al (1999) invoke the interaction of a pulsar nebula with a dark cloud to account for 3EG J 1809-2328.

At higher galactic latitudes, Wallace et al (2002) have proposed the identification of 3EG J2006-2321 with PMN J2005-2310, a flat spectrum radio quasar with a very low radio yield, while Mukherjee et al (2002) have proposed the radio galaxy NGC 6251 as the counterpart of 3EG 1621+8203. If confirmed, both identifications could have important implications for the identifications of sources at high galactic latitudes, since they provide different templates from those used so far.

## Conclusions

Although γ-ray sources are distributed all over the sky, the real puzzle rests with the identification of the low latitude, galactic ones. Currently, in fact, we lack a viable template for the bright, low latitude galactic sources. Although all the low galactic latitude sources identified so far are pulsars, their γ-ray yield seems to be too low to meet the luminosity requirements of (0.7- 16.7) $10^{35}$ erg/sec. Not even the Crab pulsar (artificially upgraded to an isotropic source) could afford to radiate so much energy.

Since radio-quiet, isolated neutron stars are even less energetic, one concludes that the real template accounting for low latitude bright galactic sources is still to be discovered.

Only systematic X-ray coverage of EGRET sources, followed by painstaking optical identification work, will provide the much needed clue(s).

In addition, the next γ-ray astronomy missions, such as AGILE and GLAST, should be prepared to take full advantage of any γ-ray source variability (as well as of their better angular resolution) to proceed towards a "γ-ray only" identification strategy.

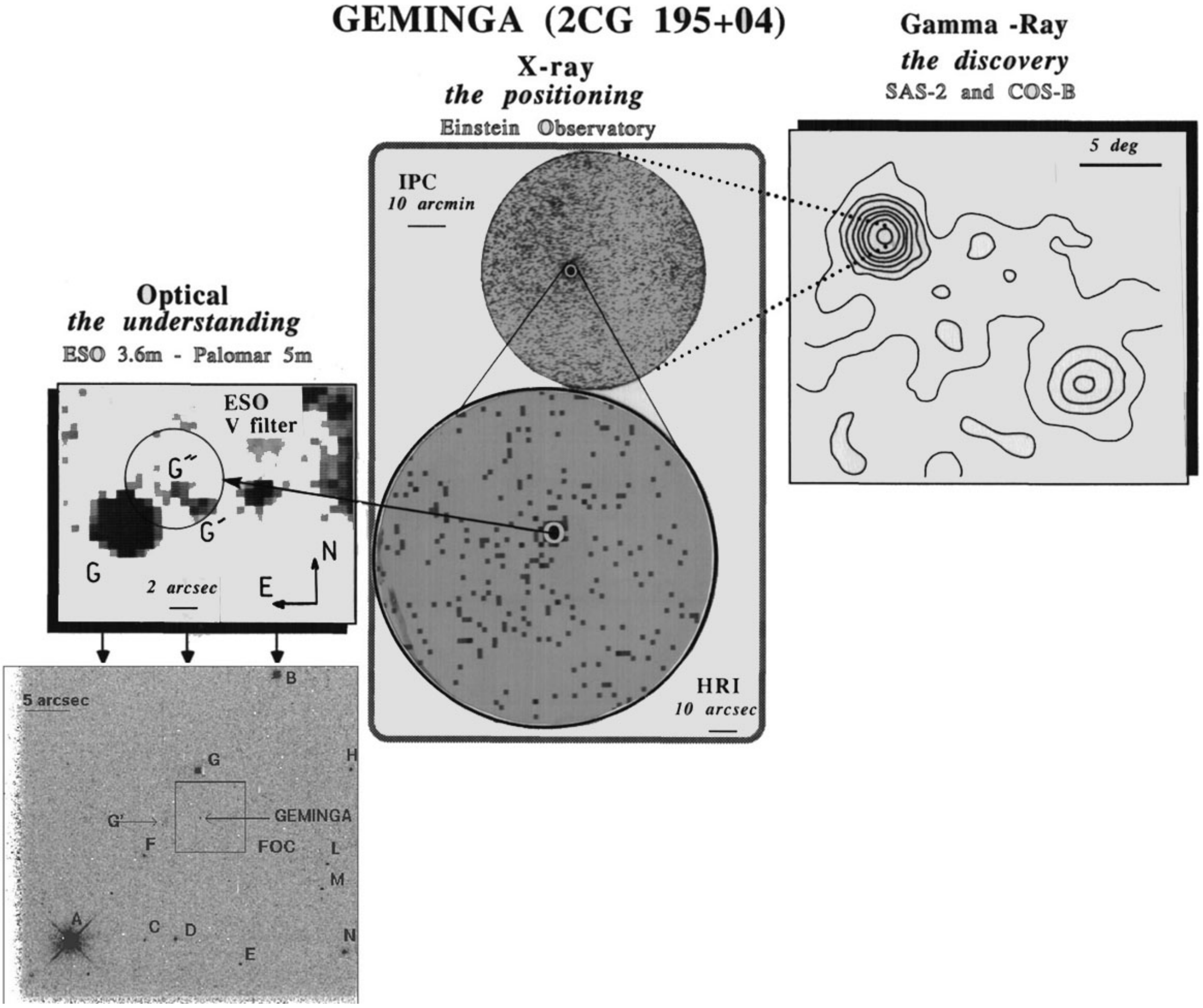

Figure 2: The identification of Geminga 1975-1996 : more than 20 years of work to secure the identification of the brightest γ-ray source in the sky with a radio-quiet pulsar. From right to left, one can follow the zoom-in approach starting from the COS-B γ-ray isophote map, proceeding through the X-ray data, gathered with the Einstein Observatory low and high resolution instruments, and ending with the optical images collected by the ESO 3.6 m telescope and the Hubble Space Telescope (for a thorough description see Bignami and Caraveo, 1996).